\documentclass[10pt,aps,prd,twocolumn,superscriptaddress,nofootinbib, floatfix]{revtex4-2}

\usepackage{graphicx}
\usepackage{amsmath,amssymb}
\usepackage{multirow}
\usepackage{adjustbox}
\usepackage{nameref}
\usepackage{hyperref}  
\usepackage{newtxtext,newtxmath}
\usepackage[normalem]{ulem}
\usepackage{dcolumn}
\usepackage{bm}
\usepackage[T1]{fontenc}
\usepackage[normalem]{ulem}
\usepackage{xcolor}

\DeclareRobustCommand{\VAN}[3]{#2}
\let\VANthebibliography\thebibliography
\def\thebibliography{\DeclareRobustCommand{\VAN}[3]{##3}\VANthebibliography}

\begin{document}

\title{Estimation of the $M_{\mathrm{TOV}}$ precision for ET, CE, and NEMO from the post-merger of BNS coalescences}

\author{Gabriela Conde-Saavedra}
\email{gabriela.saavedra@inpe.br}
\affiliation{Divisão de Astrofísica, Instituto Nacional de Pesquisas Espaciais, Av. dos Astronautas 1758, São José dos Campos, 12227-010, SP, Brazil}

\author{Odylio Denys Aguiar}
\affiliation{Divisão de Astrofísica, Instituto Nacional de Pesquisas Espaciais, Av. dos Astronautas 1758, São José dos Campos, 12227-010, SP, Brazil}

\author{H. P. de Oliveira}
\affiliation{Universidade do Estado do Rio de Janeiro, R. São Francisco Xavier, 524, CEP 20550-900, Maracanã, Rio de Janeiro, Brazil}

\author{Maximiliano Ujevic}
\affiliation{Centro de Ciências Naturais e Humanas, Universidade Federal do ABC, 09210-170, Santo André, São Paulo, Brazil}

\begin{abstract}
The detection of the gravitational waves produced after the coalescence of two neutron stars is greatly anticipated because it will be able to provide information about matter in extreme conditions, especially if the remnant turns out to go through a hypermassive or a supermassive neutron star state before collapsing into a black hole. Next-generation gravitational wave detectors such as ET, CE, and NEMO are expected to observe high-frequency gravitational wave signals, that is, the post-merger stage of the coalescence of binary neutron stars; then from these signals one can estimate the maximum mass that a spinless neutron star (M$_{\rm{TOV}}$) can have. In this paper, we investigate the problem of the determination of the M$_{\rm{TOV}}$ precision from the post-merger detected by next-generation observatories. Our results show that only under the most optimistic scenario of signal-to-noise ratio for post-merger signals ($\rm{SNR}\geq8$) and merger rate density ($250\,\rm{Gpc}^{-3}yr^{-1}$), CE achieves marginally a mass precision in the range $\delta M/2\approx 0.3-0.8\, M_{\odot}$ for a remnant mass of $2.57\,M_{\odot}.$ We clarify that this precision represents the minimum uncertainty, corresponding to the most probable value (main peak) in the final mass distribution, showing that it depends on the value of the final (remnant) mass. Therefore, based on the results obtained in this study, it will still be necessary to improve the sensitivity at high frequencies of future ground-based gravitational wave observatories if one wants to obtain greater precision in the M$_{\rm{TOV}}$ estimation. One possibility would be to improve the sensitivity in a frequency range that allows us to determine whether or not a black hole was formed in the coalescence.
\end{abstract}

\maketitle

\section{Introduction}
Numerical-relativity studies of the coalescence of binary neutron star (BNS) systems predict four possible outcomes \cite{BaiottiRezzolla2017, Sarin:2020gxb, Bernuzzi_2020, Radice_2020, Dietrich_2021}: A \textit{prompt-collapse black hole}, a \textit{hypermasive neutron star} (HMNS), a \textit{supermassive neutron star} (SMNS) -  these latter two may collapse into a black hole (BH) - or in the case of a SMNS, it may also end up as a \textit{stable neutron star}. The fate of a BNS collision depends on the characteristics of the initial neutron stars, such as their masses, spins, and equations of state (EoSs). These pre-merger parameters can be inferred from current gravitational-wave (GW) observations. There are well-developed techniques to analyze the inspiral signals, mostly because actual ground-based detectors are sensitive enough for the pre-merger stage, so the observations adjust the theoretical-numerical studies. Following the merger, the new object can be characterized by two main parameters: the total mass and the spin, as rotation plays a critical role in the mass that a remnant can retain \cite{Baumgarte2000, Weih_2017, Most:2020bba}. According to numerical studies, this post-merger (PM) stage shows better computational convergence in the frequency domain, rather than in the time domain \cite{Topolski:2023ojc}, although the strains show different features according to the EoS and the mass symmetry. Only with direct observations of the high-frequency GW from the PM stage can we have a complete understanding of the whole coalescence. That is why third-generation (3G) detectors such as the Einstein Telescope (ET) \cite{Hild:2009ns} and the Cosmic Explorer (CE) \cite{Srivastava:2022slt}; and 2.5-generation (2.5G) observatories such as Neutron-star Extreme Matter Observatory (NEMO) \cite{Ackley:2020atn, Sarin_Lasky_2022} are very promising projects for this purpose.

The GW signal emitted during the PM stage of a binary neutron star coalescence contains rich information about the properties and dynamics of the remnant object. Numerical-relativity simulations have shown that the PM GW spectrum is characterized by a dominant peak frequency, commonly referred to as $f_{\mathrm{peak}}$ or $f_2$, which is usually associated with the fundamental quadrupole ($l=2, m=2$) mode of the remnant \cite{Takami_2014, bauswein2015PM, Bernuzzi_2015}. The shape of the PM spectrum is distributed over the $(1 - 5)\;[\mathrm{kHz}]$ frequency band and the main peak is often in the middle of secondary ones, where the lower frequency peaks are thought to be generated in non-linear couplings \cite{Bauswein_2016, Chirenti_2023}. 

Furthermore, it is interesting to note that some studies have found quasi-universal relations between the frequency peak $f_{\mathrm{peak}}$ and the tidal deformability parameter $\Lambda$ extracted from the inspiral stage \cite{Bernuzzi_2014, Breschi109.064009}. This relation makes them useful tools for learning about extreme matter by linking PM signals to what we already know from the inspiral stage. In general, stiffer EoSs tend to produce broader, lower-frequency peaks due to less compact remnants, while softer EoSs lead to more compact stars and higher-frequency emissions \cite{Bauswein_2012}. 

In this paper, we use two sets of BNS simulations obtained from the Computational Relativity (CoRe) database \cite{Dietrich_2018, Gonzalez:2022mgo}. The strains were calculated at $40\,\textrm{Mpc}$ and the masses of the binary system were selected to be similar to the GW170817 event~\cite{Abbott2017_GW170817}. In the first dataset, the only varying parameter is the EoS, so the individual masses are the same for spinless neutron stars moving in a nearly zero eccentricity orbit. Instead, the second set has simulations in which the varying parameters are the masses of the initial spinless neutron stars, assuming that they both have the SLy~\cite{SLy_read2009} equation of state. With these data sets, we will explore the characteristics of the PM stage in the time domain and then we will work with their frequency domains to compare them with future detector sensitivity curves, which will be quantified by the signal-to-noise (SNR) averaged over all-sky angles and then we will explore the neutron star mass distribution of the remnant object. Previous works based on NS mass distributions inferred from Galactic binaries and GW measurements constrained the maximum mass for cold non-rotating neutron stars ($M_{\mathrm{TOV}}$), reporting maximum values around $2.1^{+0.7}_{-0.3}\,M_{\odot}$~\cite{Landry_2021} and $2.25^{+0.08}_{-0.07}\,M_{\odot}$~\cite{Fan:2023spm}. 

Our work will estimate the mass precision in  $M_{\mathrm{TOV}}$ that the next-generation detectors will obtain directly from the PM signal in BNS coalescences, considering $8$ as a reference SNR for the PM stage~\cite{Panther_2023} and $7.6-250\,\textrm{Gpc}^{-3}\textrm{yr}^{-1}$~\cite{LIGOScientific:2025newR0} as the BNS merger rate. Here, we point out that we are optimistically assuming that a $\rm{SNR}_{ref}\geq8$ for the post-merger signal will be sufficient for 3G interferometers to observe the formation of a black hole even in the case of a delayed collapse, since they aim to have extended sensitivities to the highest possible frequencies to detect the quasi-normal modes of a black hole, even if formed late. 

In this article, Sec.~\ref{sec:bnssimulations} describes the properties of the BNS simulations obtained from the CoRe database to perform our calculations. In addition, we outline the method to study the frequency domain of the simulations, as well as the matched-filtering technique to compare the BNS frequency signal with the sensitivity curves of the next-generation detectors. In Sec.~\ref{sec:nsmasslimits} we discuss the neutron star mass distributions in binary systems with the aim of estimating the final mass distribution and calculating the mass precision of $M_{\mathrm{TOV}}$. Finally, the results and perspectives are summarized in Sec.~\ref{sec:concperspec}.

\section{\label{sec:bnssimulations}Binary Neutron Star Merger Simulations}
In this section, we explore binary neutron star coalescence through numerical simulations obtained from the CoRe (Computational Relativity) database~\cite{Gonzalez:2022mgo}. These simulations provide gravitational wave strain data for the full coalescence process. We begin by describing the physical parameters of the strain waveforms $h_{22}$ in the time domain, followed by an analysis of their amplitude spectral density (ASD) with focus on the frequencies of the PM signal. Finally, we study the detectability of these signals by computing the signal-to-noise ratio (SNR), averaged over all-sky angles, using the matched filtering technique and considering the sensitivity curve of three future gravitational wave detectors.

\subsection{\label{sec:coresims}CoRe Database}
The time-domain strain simulations taken from the CoRe database (\cite{Gonzalez:2022mgo}) are based on two types of code: BAM (Bi-functional Adaptive Mesh \cite{Brugmann_2004}) and THC (Templated-Hydrodynamics Code \cite{Radice_2012THC}). Both are developed in a general relativistic hydrodynamical framework, but at the moment, the simulations with neutrino cooling presented in the CoRe database were done with the THC code.

The strain of a gravitational wave emitted during the coalescence of a binary compact object can be written in terms of two polarizations $h_{+}$ and $h_{\times}$, which are the modes of oscillation around the direction of propagation \cite{Krolak_2023}, 
\begin{equation}
h_{+}-ih_{\times}=\frac{M}{r}\sum_{l=2}^{\infty}\sum_{m=-l}^{l}H_{lm}(t)\; ^{-2}Y_{lm}(\theta, \phi),
\end{equation}

\noindent where $M$ is the total mass of the binary system; $r$ is the distance from Earth to the binary system; $H_{lm}(t)$ represents the time-dependent amplitudes of the gravitational wave signal projected onto the spin-weighted spherical harmonic modes, obtained from numerical relativity simulations or from approximate methods used to solve the Einstein field equations, such as Post-Newtonian (PN) or Effective-One-Body (EOB) formalisms; ${}^{-2}Y_{lm}(\theta, \phi)$ are the spin-weighted spherical harmonics of spin weight $-2$, which account for the angular distribution of the gravitational wave radiation; and $\theta$ and $\phi$ are the polar and azimuthal angles, respectively, that define the orientation of the system with respect to the Earth.

In compact binary coalescences, the dominant contribution to gravitational radiation comes from the quadrupole mode with $l=m= 2$. This mode $h_{22}$ represents the highest intensity emission and governs the characteristic ``chirp'' signal that is observed by the detectors. Higher-order modes $(l>2)$ can also contribute, particularly in systems with high mass ratios or significant orbital eccentricity, but their amplitudes are weaker compared to the $l=m=2$ component.

\begin{figure}
\includegraphics[width=\columnwidth]{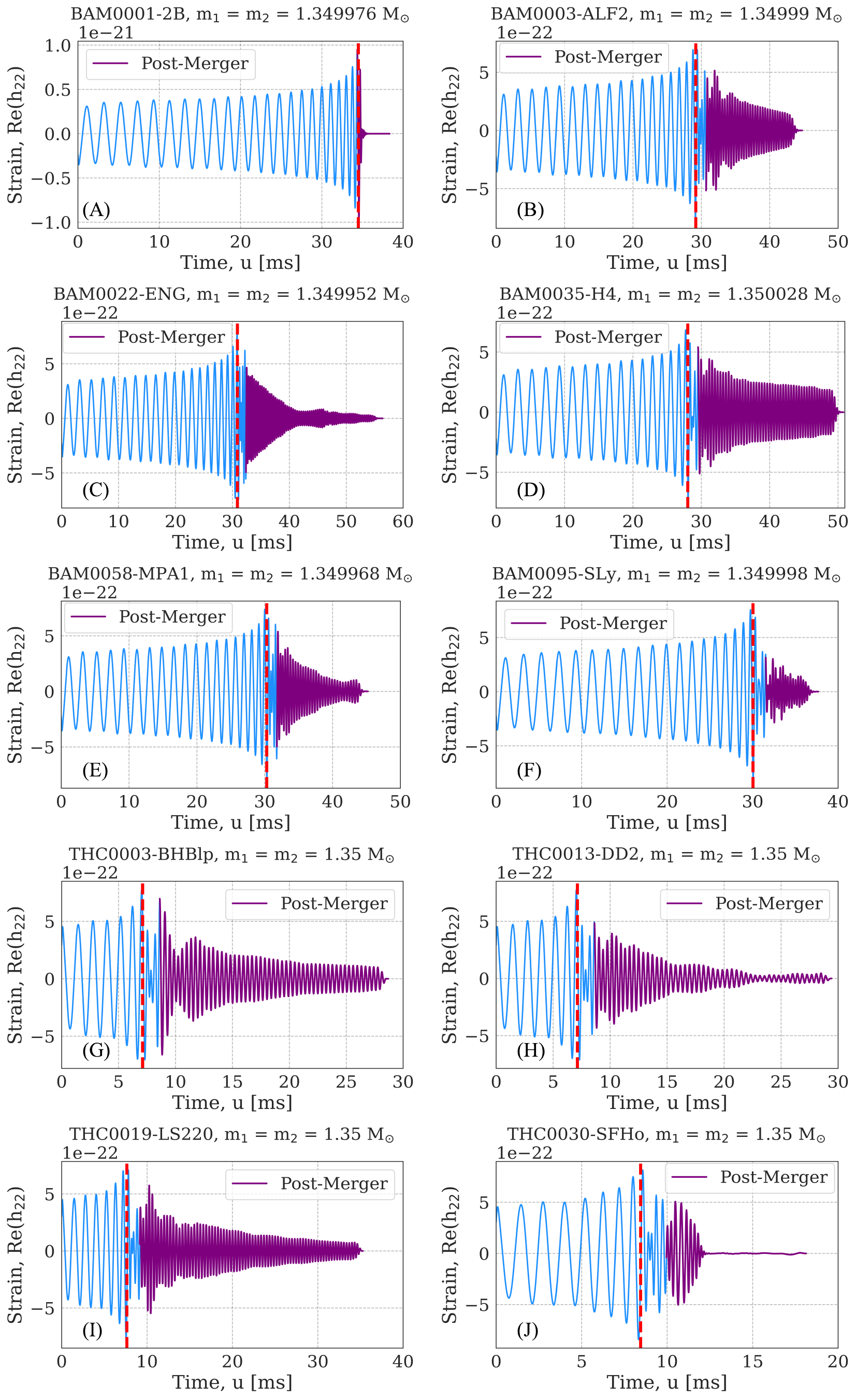}
\caption{\label{fig:TScombinedDataset1} Strain time-series of the Dataset 1. The simulations were extracted at a distance $40\,\textrm{Mpc}$. They have components of equal mass or mass ratio equal to one. Each simulation has a different equation of state. The PM stage is the purple line and the vertical dashed line represents the amplitude (Eq.~\ref{eq:wveformamplitude}) at its maximum value, defining the end of the inspiral stage.}
\end{figure}

For Dataset 1, we selected ten BNS systems with total mass of $M_T=m_1+m_2=2.70\;M_{\odot}$, mass ratio $q=m_1/m_2=1$, and different EoS, see Tab.~\ref{tab:Dataset1} for more details and see Fig.~\ref{fig:TScombinedDataset1} to visualize the shape of the waveforms. The individual stars of the simulations are spinless and their characteristics are consistent with the component masses of GW170817, including the Chirp Mass~\cite{Maggiore:2007ulw} defined as
\begin{equation}
\mathcal{M}_{\rm{Chirp}}=\frac{(m_{1}m_{2})^{3/5}}{M_{\rm{T}}^{1/5}}.
\end{equation}

\noindent For Dataset 1 we have considered distinct EoS: 2B, ALF2, ENG, H4, MPA1 and SLy~\cite{SLy_read2009}; BHBlp~\cite{Banik_2014BHBlp}; DD2~\cite{Typel_2010DD2}; LS220~\cite{1991NuPhA_LS220}; and SFHo~\cite{SFH2013ApJ77417S}. See Tab.~\ref{tab:Dataset1} to identify the names of the selected simulations, which were chosen to represent diverse nuclear matter properties, including hadronic, hybrid, and models that include the presence of exotic phases, such as quark matter in BHBlp. 

\begin{table*}
\centering
\caption{Dataset 1: The simulations have the same gravitational mass components during the inspiral, $m_{1}=m_{2}=1.35\;M_{\odot}$, with $q=1$ and $\mathcal{M}_{\rm{Chirp}}=1.175\;~M_{\odot}$, but different EoS for which we choose a variety of models: hadronic, hybrid, and models that include exotic matter. From left to right, the columns represent the simulation name, the EoS, the main peak frequency in the ASD of the PM phase, the main frequency peak in the characteristic strain spectra, and the last three columns correspond to the averaged SNR and its uncertainty for ET, CE and NEMO detectors.}
\begin{tabular}{ l c c c c c c }
\hline \hline
Name & EoS & PM ASD $f_{2}$ & $h_c$ main peak & ET SNR & CE SNR & NEMO SNR \\
  &   & [kHz] & [kHz] &   &   &   \\ \hline
BAM0001 & 2B    & 3.625 & 3.883 & $3.39\pm2.30$  & $6.22\pm3.55$ & $3.89\pm2.24$  \\
BAM0003 & ALF2  & 2.704 & 2.704 & $6.68\pm3.47$ & $11.75\pm6.09$ & $9.69\pm5.01$ \\
BAM0022 & ENG   & 2.871 & 2.871 & $4.73\pm2.56$ & $8.27\pm4.48$ & $6.57\pm3.55$ \\
BAM0035 & H4    & 2.488 & 2.488 & $8.69\pm4.68$ & $15.3\pm8.23$ & $12.85\pm6.90$ \\
BAM0058 & MPA1  & 2.609 & 2.609 & $5.24\pm2.87$ & $9.15\pm5.01$ & $7.50\pm4.09$ \\
BAM0095 & SLy   & 3.405 & 3.405 & $2.55\pm1.39$  & $4.30\pm2.34$  & $3.15\pm1.72$  \\
THC0003 & BHBlp & 2.637 & 2.637 & $7.57\pm4.09$ & $13.22\pm7.12$ & $10.78\pm5.77$ \\
THC0013 & DD2   & 2.498 & 2.498 & $5.81\pm3.09$ & $10.14\pm5.39$ & $8.39\pm4.45$ \\
THC0019 & LS220 & 2.868 & 2.868 & $6.86\pm3.55$ & $12.06\pm6.24$ & $9.73\pm5.04$ \\
THC0030 & SFHo  & 3.545 & 4.532 & $2.90\pm1.54$ & $4.99\pm2.65$ & $3.68\pm1.96$  \\
\hline \hline
\end{tabular}
\label{tab:Dataset1}
\end{table*}

\begin{figure}
\includegraphics[width=\columnwidth]{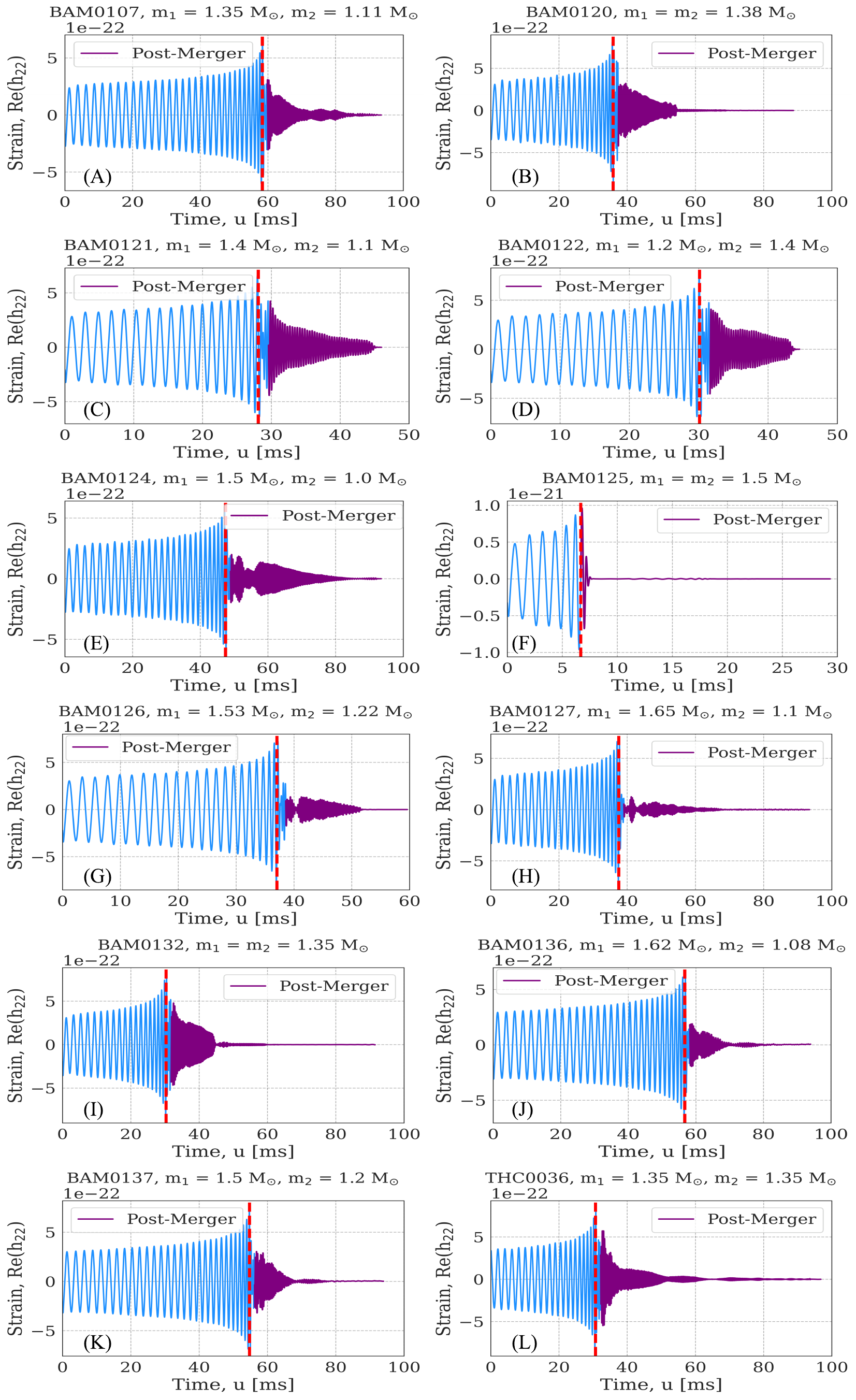}
\caption{\label{fig:TScombinedDataset2} Strain time-series of the Dataset 2 located at $40\,\textrm{Mpc}$. The equation of state is fixed for all simulations: SLy. The mass ratio is not necessarily equal to one. The PM stage is marked as a purple line and the vertical dashed line represents the amplitude (Eq.~\ref{eq:wveformamplitude}) at its maximum value, defining the end of the inspiral stage.}
\end{figure}

Dataset 2, Tab.~\ref{tab:Dataset2}, see also Fig.~\ref{fig:TScombinedDataset2}, contains twelve BNS simulations with the same SLy EoS, but the mass ratio is not necessarily equal to one. Also, in this case, the individual stars are spinless. The range of individual masses is between $(1.00 - 1.65)\;M_{\odot}$, the total mass is between $(2.44 - 3.00)\;M_{\odot}$, and the Chirp Mass is between $(1.08 - 1.31)\;M_{\odot}$. In addition, the orbit also has nearly zero eccentricity. 

In Fig.~\ref{fig:TScombinedDataset1} and Fig.~\ref{fig:TScombinedDataset2} the merger time is marked with a dashed vertical line. This moment was obtained by searching for the maximum value of the waveform's amplitude, 
\begin{equation}
A=\sqrt{{\rm Re}^2(h_{22})+{\rm Im}^2(h_{22})}=\sqrt{h_{22,+}^{2}+h_{22,\times}^{2}}.
\label{eq:wveformamplitude}
\end{equation}

\noindent The interval of $1.5\;\mathrm{ms}$ just after the inspiral phase characterizes the merger, from which the post-merger stage starts, which is the object of interest in this work and is marked in purple. This time interval is not completely arbitrary, and it is
based on studies about prompt black hole collapse times. Numerical studies, using 
the method of the minimum lapse, suggest that the time for a prompt black hole
collapse is approximately $\leq 2$~ms \cite{kolsch2022investigating}, a similar
value is also considered in \cite{agathos2020inferring}. As we can observe in both figures, the shape of the PM strain varies in both data sets, showing the different characteristics of the remnant object.

\begin{table*}
\caption{Dataset 2: Simulations with different $q=m_1/m_2$ but the same SLy EoS. From left to right, the columns represent the simulation name, the gravitational mass of star 1, the gravitational mass of star 2, the total gravitational mass of the system, the chirp mass, the mass ratio, the main peak frequency in the ASD of the PM phase, the main frequency peak in the characteristic strain spectra, and the last three columns correspond to the averaged SNR and its uncertainty for ET, CE and NEMO detectors.}
\begin{tabular}{lcccccccccc}
\hline \hline
Name & $m_{1}$ & $m_{2}$ & $M_{\rm{T}}$ & $\mathcal{M}_{\rm{Chirp}}$ & $q$ & PM ASD $f_{2}$ & $h_c$ main peak & {ET SNR} & {CE SNR} & {NEMO SNR} \\
   &  [$M_{\odot}$] & [$M_{\odot}$] & [$M_{\odot}$] & [$M_{\odot}$] & & [kHz] & [kHz] &   &   &   \\ \hline
BAM0107 & 1.35 & 1.11 & 2.46 & 1.11 & 1.22 & 2.708 & 2.708 & $3.44\pm1.85$  & $6.03\pm3.23$ & $4.94\pm2.64$ \\
BAM0120 & 1.38 & 1.38 & 2.76 & 1.20 & 1.00 & 3.428 & 3.428 & $4.14\pm2.23$ & $7.15\pm3.84$ & $5.09\pm2.73$ \\
BAM0121 & 1.40 & 1.10 & 2.44 & 1.08 & 1.22 & 2.808 & 2.808 & $4.42\pm2.28$  & $7.71\pm3.98$ & $6.24\pm3.22$ \\
BAM0122 & 1.40 & 1.20 & 2.60 & 1.13 & 1.17 & 3.016 & 3.016 & $4.25\pm2.30$  & $7.41\pm3.99$ & $5.69\pm3.06$ \\
BAM0124 & 1.50 & 1.00 & 2.50 & 1.06 & 1.50 & 2.921 & 2.921 & $3.36\pm1.80$  & $5.89\pm3.15$ & $4.76\pm2.55$  \\
BAM0125 & 1.50 & 1.50 & 3.00 & 1.31 & 1.00 & 2.513 & 2.778 & $6.17\pm3.34$ & $9.67\pm5.20$ & $6.16\pm3.33$  \\
BAM0126 & 1.53 & 1.22 & 2.75 & 1.19 & 1.25 & 3.329 & 3.329 & $1.86\pm1.03$  & $3.15\pm1.73$  & $2.21\pm1.20$  \\
BAM0127 & 1.65 & 1.10 & 2.75 & 1.17 & 1.50 & 3.438 & 3.447 & $1.33\pm0.74$  & $2.30\pm1.28$  & $1.66\pm0.92$  \\
BAM0132 & 1.35 & 1.35 & 2.70 & 1.18 & 1.00 & 3.317 & 3.324 & $4.97\pm2.75$ & $8.63\pm4.77$ & $6.28\pm3.46$ \\
BAM0136 & 1.62 & 1.08 & 2.70 & 1.15 & 1.50 & 2.993 & 2.990 & $2.18\pm1.16$  & $3.77\pm2.00$  & $2.94\pm1.55$  \\
BAM0137 & 1.50 & 1.20 & 2.70 & 1.17 & 1.25 & 3.129 & 3.131 & $3.45\pm1.77$  & $5.98\pm3.06$ & $4.56\pm2.33$ \\
THC0036 & 1.35 & 1.35 & 2.70 & 1.18 & 1.00 & 3.409 & 3.409 & $4.72\pm2.47$ & $8.22\pm4.29$ & $6.06\pm3.17$ \\
\hline \hline
\end{tabular}
\label{tab:Dataset2}
\end{table*}

In the next subsection, we will look at the PM signals in the frequency domain to identify possible features in the spectra shape that could tell us about the remnant properties. By comparing the results from both datasets, we will see how the EoS and mass ratio affect the main frequencies of the spectra.

\subsection{\label{sec:freqdomain}Amplitude Spectral Density}
After obtaining the time-domain waveform for the PM stage in each simulation, we calculate the Fast Fourier Transformation (FFT) amplitude spectrum. The waveform is a time series $h(t)$ and the Fourier transformation is defined as~\cite{Moore_2014}
\begin{equation}
    \tilde{h}(f)=\int_{-\infty}^{\infty} \mathrm{d}t\; h(t)\; \exp{(-2\pi \mathrm{i} ft)}.
	\label{eq:ffteq}
\end{equation}

Subsequently, the Fast Fourier Transform over a frequency resolution will give the Power Spectral Density (PSD), defined as~\cite{Moore_2014}
\begin{equation}
    {\rm PSD}(f)=\frac{1}{Nf_{s}}|\tilde{h}(f)|^{2},
	\label{eq:psdeq}
\end{equation}

\noindent where $\Tilde{h}(f)$ is the FFT frequency value, $Nf_{s}$ is the number of elements times the sampling frequency $f_{s}$. The sampling frequency is the inverse of the time difference between the consecutive data points $f_{s}=1/[t(n)-t(n-1)]$, where $n=1,...,N$. Finally, the Amplitude Spectral Density (ASD) is defined as the square root of PSD~\cite{Moore_2014}
\begin{equation}
    {\rm ASD}(f)=\sqrt{{\rm PSD}(f)}.
	\label{eq:asdeq}
\end{equation}

Generally, to compare the sensitivity of a detector and the GW signal, we use the characteristic strain $h_c$, which can be defined in terms of ASD and PSD as~\cite{Moore_2014}
\begin{equation}
    h_c(f)=\rm{ASD}\sqrt{f}=\sqrt{f\,\rm{PSD}}.
\end{equation}

In the case of the GW time series, the strain is dimensionless. The units of PSD and ASD spectra are [$\rm{Hz}^{-1}$], [$\rm{Hz}^{-1/2}$], so $h_c$ is dimensionless. Once we get the ASD spectra, we identify the main peak frequencies in each spectrum. 

Table~\ref{tab:Dataset1} presents the main peak frequencies in the amplitude spectral density (ASD) of the PM stage for Dataset 1. There is a variation in the PM frequency $f_2$ depending on the equation of state of the inspiral binary system. Stiffer EoSs, such as H4 and DD2, generally lead to lower peak frequencies ($2.488\;\rm{kHz}$ and $2.498\;\rm{kHz}$, respectively), which is related to the formation of less compact remnants. In contrast, softer EoSs such as SFHo and SLy are related to a remnant with higher frequencies ($3.545\;\rm{kHz}$ and $3.405\;\rm{kHz}$, respectively), reflecting its more compact nature. 

Table ~\ref{tab:Dataset2} displays the main peak frequencies $f_2$ in the ASD spectra for PM remnants of Dataset 2. The results suggest that the peak frequency increases with an increasing total mass, reflecting the effect of the high compactness of more massive remnants. For example, systems such as BAM0107 $(2.46\;M_{\odot})$ and BAM0121 $(2.44\;M_\odot)$ exhibit peak frequencies around $(2.7 - 2.8)\;\rm{kHz}$, while higher-mass systems such as BAM0120 $(2.76\;M_{\odot})$ and THC0036 $(2.70\;M_{\odot})$ show frequencies above $3.4\;\rm{kHz}$. For the case of BAM0125 with the highest mass in the table $(3.00\;M_{\odot)}$, it presents a lower peak frequency of $2.5\;\rm{kHz}$, which may suggest the formation of a prompt-collapse black hole. Furthermore, for systems with the same total mass $(2.70\;M_{\odot})$, the values of $f_2$ show some dispersion (ranging from $2.993\;\rm{kHz}$ to $3.409\;\rm{kHz}$), possibly reflecting the influence of the mass ratio on the dynamics of BNS coalescence. These results reinforce the sensitivity of the PM spectral features to the binary mass configuration, providing further opportunities to constrain neutron star properties through gravitational wave observations. In summary, for all simulations, the PM main frequencies fall between $(2.488 - 3.625)\; \rm{kHz}$.  

\begin{figure}
\includegraphics[width=\columnwidth]{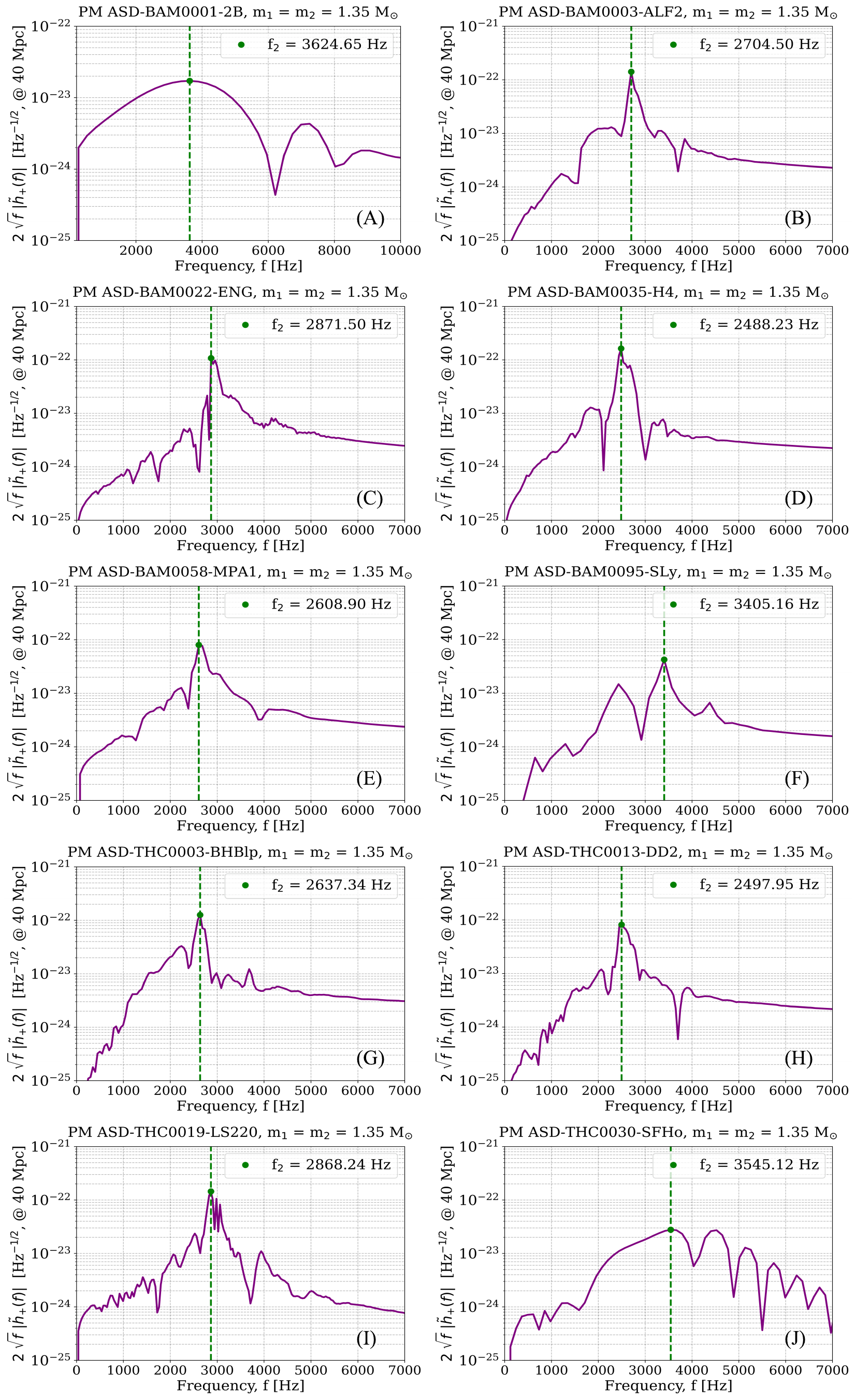}
\caption{\label{fig:combinedPMASD1} Amplitude Spectral Density for Dataset 1. They are $q=1$ simulations that differ in the EoS. The main peaks in the ASD spectra are between $2.488-3.625\;\textrm{kHz}$, being the lowest value for the stiff EoS H4 and the highest value for the soft EoS 2B. In (A) we see that BAM0001 is particularly different from the rest of spectra because its time-domain has the shape of a prompt-collapse black hole, then it is possible to find additional peaks at $f\sim 6-8\,\textrm{kHz}$.}
\end{figure}

\begin{figure}
\includegraphics[width=\columnwidth]{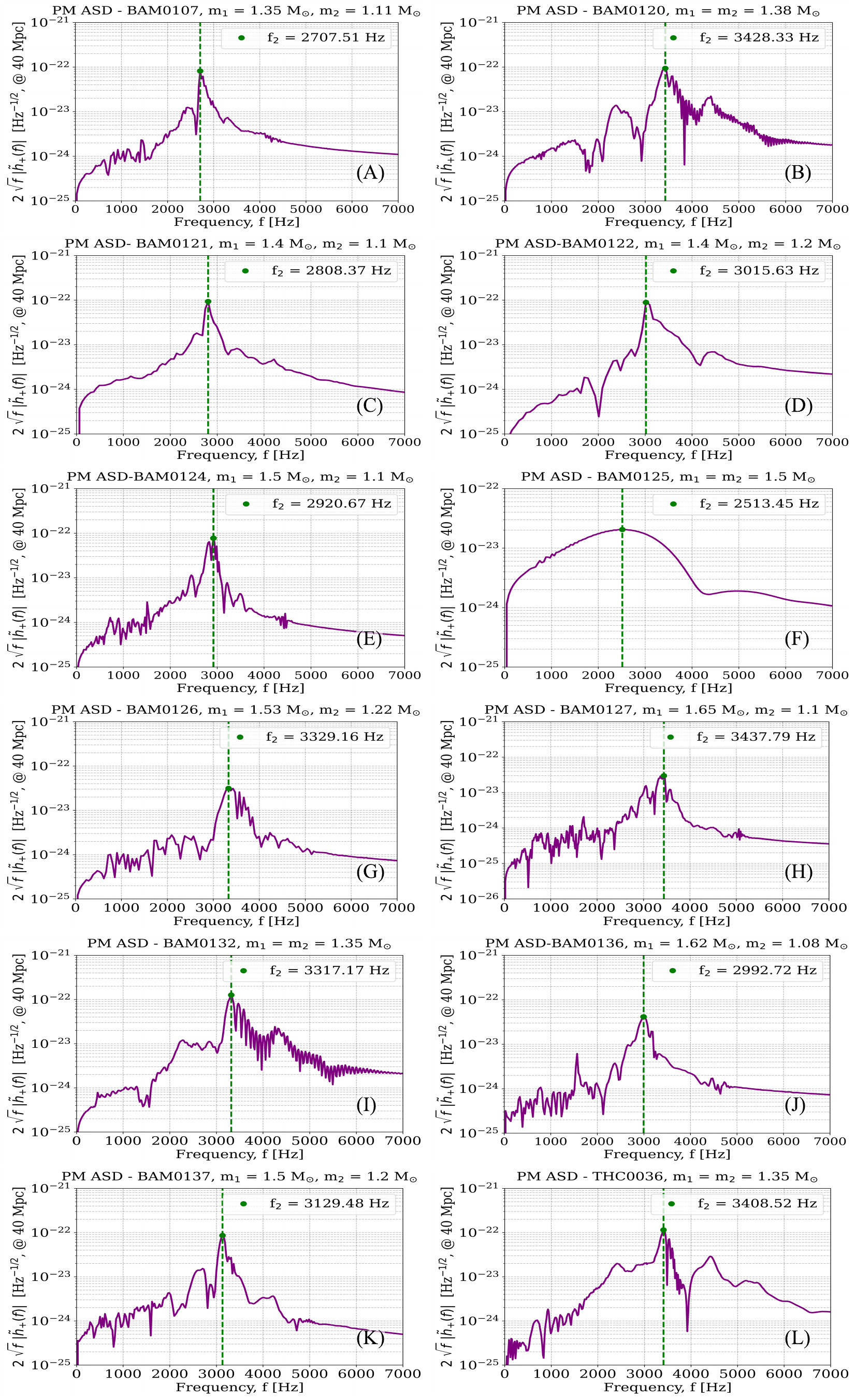}
\caption{\label{fig:combinedPMASD2} Amplitude Spectral Density for Dataset 2. They have the same SLy EoS, but their mass ratio is not necessarily equal to $1$. The main peak frequencies in the ASD spectra are between $2.513-3.438\,\textrm{kHz}$. (F) shows a single peak at $2.513\,\textrm{kHz}$ from a, most probably, hypermassive neutron star (HMNS). In general, the spectra are less smooth compared to the spectra of data set 1, probably because of the mass-asymmetry and therefore, the irregularities in the oscillation of the remnant.}
\end{figure}

The shapes of the spectra, see Fig.~\ref{fig:combinedPMASD1} and Fig.~\ref{fig:combinedPMASD2}, look smoother in Dataset 1 than in Dataset 2. Probably because the mass-asymmetry in Dataset 2 produces more irregularity in the oscillation of the PM stage. It is interesting to note that independently of the inspiral mass values, BAM0001 and BAM0125 show a broad peak around $3624.65\,\rm{kHz}$ and $2513.45\,\rm{kHz}$, respectively. The BNS with EoS 2B and total mass $2.70\;M_{\odot}$ forms a remnant with $f_{2}=3.625\;\rm{kHz}$. However, BNS with EoS SLy and total mass $3.00\;M_{\odot}$ forms a remnant with a lower PM frequency $f_{2}=2.513\;\rm{kHz}$. See Fig.~\ref{fig:combinedPMASD1}-(A) and Fig~\ref{fig:combinedPMASD2}-(F), respectively. 

As we have seen, the different shapes in the spectra vary according to the characteristics of the binary components; then these spectra may hold important information about the post-merger remnant. In the next subsection, we evaluate the detectability of high-frequency band GW detectors using signal-to-noise and matched filtering techniques.

\subsection{\label{sec:snr}Signal-to-Noise Ratio and Matched Filtering}
Gravitational wave detectors are very sensitive to noise, which may hide the true GW signal. In order to distinguish real events from noise, we use the signal-to-noise ratio to quantify the strength of the signal relative to the background noise. Then, with the matched-filtering technique, we maximize the SNR by correlating the signals that detectors would observe with theoretical templates extracted from the simulations.

Based on this, we compare these PM frequency-domain signals with the sensitivity curves of three detectors: ET-C \cite{ETsensitivity2024}, CE \cite{CEsensitivity2024}, and NEMO \cite{Ackley:2020atn}, as we have previously done for a different set of simulations \cite{10.3389/fspas.2026.1721816} from CoRe database. The response function of a gravitational wave detector tells us how the detector observes the gravitational wave signal, considering the two polarizations $h_{+}$ and $h_{\times}$ of the post-merger stage combined with the antenna pattern function $F_{+}(\theta, \phi, \psi)$ and $F_{\times}(\theta,\phi,\psi)$, which depends on the orientation of the GW source relative to the detector. The angles that describe the direction of the gravitational wave source in the sky are the azimuthal ($\theta$) and polar ($\phi$) angles. The third angle, $\psi$, represents the angle of polarization of the wave or how the shape of the wave is oriented relative to the detector (whether it is $h_{+}$ or $h_{\times}$). Additionally, for binary systems, the factors $(1 + \cos^{2} \iota)/2$ and $\cos \iota$ depend on the inclination angle $\iota$ between the angular momentum of the binary and the line of sight of the detector. Therefore, the total measured signal is \cite{Creighton2011}
\begin{equation}
h(t)=\frac{1}{2}(1+\cos^{2}\iota) F_{+}(\theta,\phi,\psi)h_{+}(t) + \cos \iota \,F_{\times}(\theta, \phi, \psi)h_{\times}(t).
\label{eq:h_T}
\end{equation}

The output response of a detector $O(t)$ includes the contribution of instrumental noise $n(t)$ and the total measured signal $h(t)$, so
\begin{equation}
O(t)=n(t)+h(t).
\label{eq:Output}
\end{equation}

In general, the SNR is defined as follows~\cite{mendonca_2023}
\begin{equation}
\rho=2 \frac{{\rm Re}\{\int_0^\infty \frac{{\rm d}f}{S_{n}} \tilde{h}(f) \tilde{v}^{*}(f)\}}{\sqrt{\int_0^\infty \frac{{\rm d}f}{S_{n}} |\tilde{v}(f)|^{2}}},    
\end{equation}

\noindent where the quantity $S_n\equiv S_n(f)$ is the one-sided power spectral density (PSD) of the detector noise. It has units of $\rm Hz^{-1}$ and defines the detector sensitivity curve. The asterisk in $\tilde{v}^{*}(f)$ denotes the complex conjugate of the Fourier transform of the template. When we assume that the signal matches perfectly with the template, the SNR is maximized by $\tilde{v}(f)=\tilde{h}(f)$, so the optimal SNR represents the maximum possible signal strength that can be detected from the noise and is expressed as~\cite{mendonca_2023}
\begin{equation}
\rho_{\mathrm{opt}}=\sqrt{\langle h|h\rangle}, 
\end{equation}

\noindent where the scalar product is defined as~\cite{mendonca_2023}

\begin{equation}
    \langle A|B \rangle \equiv 4\; \mathrm{Re} \left\{ \int_0^\infty \frac{df}{S_n(f)}\tilde{A}(f)\tilde{B}^{*}(f) \right\}.
\end{equation}

Instead, when the detected signal includes noise, we get the actual SNR. Here, we take the Fourier Transform of the output signal, $\tilde{O}(f)$, obtained from Eq.~\ref{eq:Output}, and compare it with the Fourier Transform of the template PM signal, $\tilde{h}_{T}$, derived from Eq.~\ref{eq:h_T}). Also, we need to use a time-shifting term ($\exp^{-2i \pi ft_{c}}$) in order to have both signals aligned in time. Since we do not know the exact value of the coalescence time or the arrival time of the signal at the detector, we scan possible values of $t_{c}$ to maximize the match between the output signal and the template. The expression for this case is the following \cite{mendonca_2023}
\begin{equation}
\rho = 2\frac{{\rm Re}\{ \int_0^\infty \frac{{\rm d}f}{S_{n}} \tilde{O}(f) \tilde{h}^{*}_{T}(f) \exp^{-2i \pi ft_{c} } \}}{ \sqrt{\int_0^\infty \frac{{\rm d}f}{S_{n}(f)} |\tilde{h}_{T}(f)|^{2} }}. 
\end{equation}

Since $h_T$ depends on the mentioned four angles, then we average the SNR over them, each following the distributions: $P(\theta)\propto\sin \theta $ (uniform on the sphere), $P(\iota)\propto \sin \iota$ (uniform), $P(\phi)\propto \rm{const}$, $P(\psi)\propto \rm{const}$ \cite{Creighton2011, Maggiore:2007ulw}. This average is approximated by a Monte Carlo sum
\begin{equation}
    \bar{\rho}\approx \frac{1}{N_{s}}\sum _{i=1}^{N_{s}}\rho(\theta_{i}, \phi_{i}, \psi _{i}, \iota _{i}),
\end{equation}

\noindent where $\theta_i,\,\phi_i,\,\psi_i,\,\iota_i$ are random samples according to their distributions. This ensures that the SNR is not calculated for a particular case, which might be considered a strong assumption.

Table~\ref{tab:Dataset1} shows the value of the main peak in the characteristic strain spectra and the SNRs of the three detectors for Dataset 1. The mean SNR and its standard deviation for each detector is $5.44 \pm 2.95$ (ET), $9.54 \pm 5.11$ (CE) and $7.62 \pm 4.07$ (NEMO). We identify the lower values: SNR$_{\mathrm{low}}=[2.55\pm1.39, 4.30\pm2.34, 3.15\pm1.72]$ for ET, CE and NEMO, respectively, and the corresponding highest values: SNR$_{\mathrm{high}}=[8.69\pm4.68, 15.30\pm8.23, 12.85\pm6.90]$. It is interesting to observe that the lowest SNR belongs to a simulation that uses SLy (BAM0095), and the highest values are for BAM0035, which uses H4. Based on the estimates obtained for GW170817, these EoSs are actually the best favored. Observe also that H4 EoS (BAM0035) produces a remnant with $\mathrm{SNR}=15.30\pm8.23$ and a peak frequency $h_{c,\mathrm{peak}}=2.488\;\rm{kHz}$ in the CE detector. However, SLy EoS (BAM0095) yields a remnant with the lowest detection at higher frequencies such as $3.405\;\rm{kHz}$, falling in the less sensitive region of the detector frequency range.  

We can estimate the relative performance improvement between the detectors by comparing the central values of the SNRs. For the highest SNR case (H4 EoS - BAM0035), CE's SNR of $15.30$ represents a $(15.30-8.69)/8.69 \times 100\%=76\%$ improvement over ET and a $19\%$ improvement over NEMO. Even in the lowest SNR configuration (SLy EoS - BAM0095), CE maintains a $69\%$ advantage over ET ($4.30$ vs. $2.55$) and a $37\%$ advantage over NEMO ($4.30$ vs. $3.15$). In the end, the detection capacity of CE is more pronounced against ET and NEMO in all equations of state. Therefore, the Cosmic Explorer seems to achieve the best SNRs for all PM binary neutron star simulations extracted at $40\;\mathrm{Mpc}$, with values ranging from $4.30\pm2.34$ (SLy EoS) to $15.30\pm8.23$ (H4 EoS). 

\begin{figure}
  \centering
  \includegraphics[width=\linewidth]{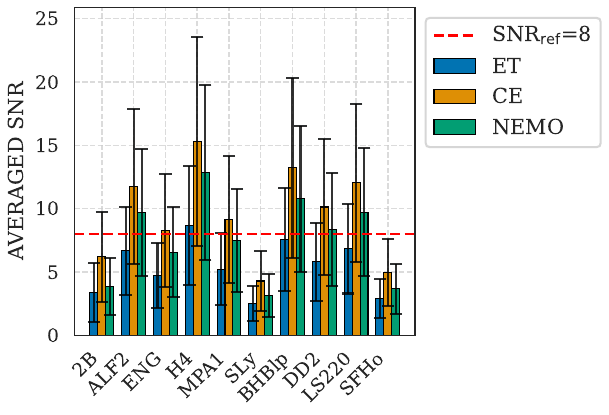}
  \vspace{0.5em} 
  \includegraphics[width=\linewidth]{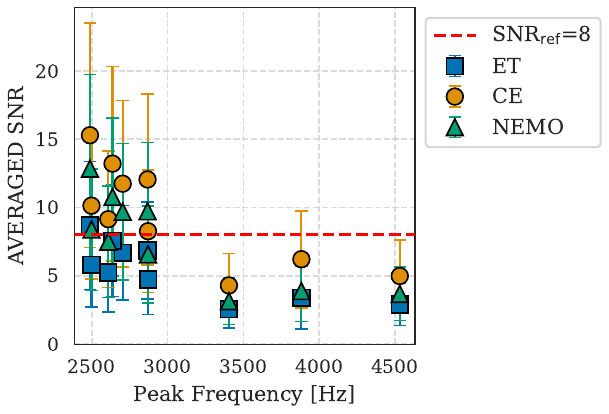}
  \caption{Detector sensitivity for Dataset 1. {\it Upper panel:} Averaged SNR as a function of the EoS for the three detectors compared with the value $\mathrm{SNR}=8$. The H4 EoS provides a remnant with the highest values of SNR. {\it Lower panel:} SNR as a function of the main peak frequency in the PM $h_c$. High-frequency region ($>3300\,\rm{Hz}$) shows the lowest central values of SNR for the three detectors.} \label{fig:detectorSNRs}
\end{figure}

Fig.~\ref{fig:detectorSNRs} shows the dependence of the SNR response of each detector according to the EoS (upper panel) and to the main peak frequency of the PM stage (lower panel). We observe that some simulations are below the limit value of $\mathrm{SNR}=8$, and belong to high frequency values, just where the three detectors are less sensitive ($\sim 3.3\;\mathrm{kHz}$).  

Now, we analyze binary neutron star mergers with varying mass ratios $(q=1.0 - 1.5)$ and fixed EoS (SLy). The results for Dataset 2 are shown in Table~\ref{tab:Dataset2}, so we observe SNR$_{\mathrm{low}}=[1.33\pm0.74, 2.30\pm1.28, 1.66\pm0.92]$ for ET, CE and NEMO, corresponding to BAM0127 ($q=1.50$). However, the SNR$_{\mathrm{high}}=[6.17\pm3.34, 9.67\pm5.20, 6.28\pm3.46]$ corresponds to different simulations: BAM0125 (ET, CE) and BAM0132 (NEMO) with mass ratio $q=1.00$. Besides, the mean values for each detector are $13.70\pm 5.07$ (CE), $7.79 \pm 3.14$ (ET) and $9.63\pm 3.40$ (NEMO). 

\begin{figure}
\includegraphics[width=\columnwidth]{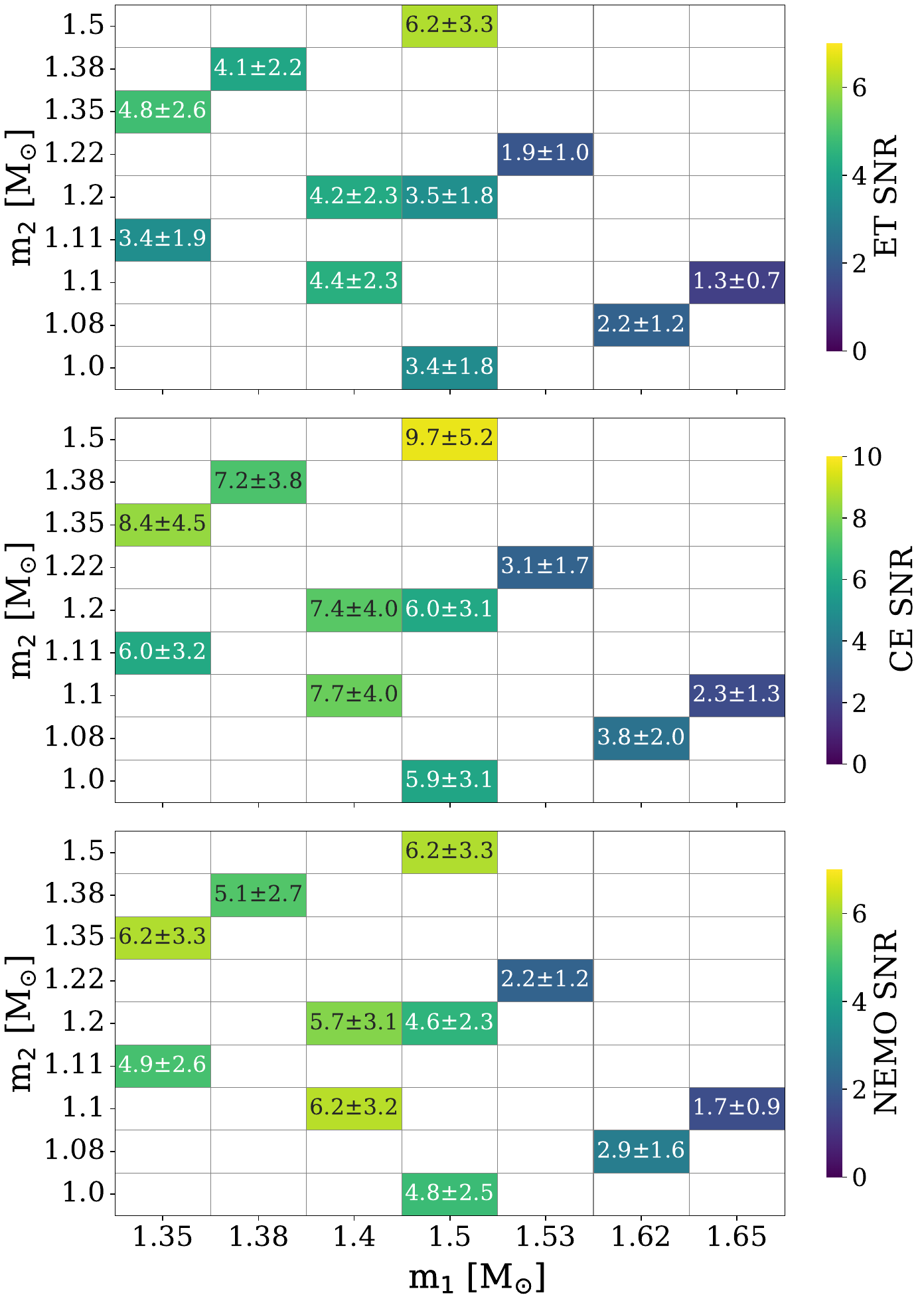}
\caption{\label{fig:heatmapsnr} Post-merger averaged SNR for Dataset 2. These plots show the central values of SNR for different detectors and ($m_1,m_2$) combinations. The yellow section of the colormap indicates the highest values for each detector, then CE has the highest values when compared to NEMO and ET.}
\end{figure}

Since Dataset 2 has more variables compared to Dataset 1, we use the heatmap representation (see Fig.~\ref{fig:heatmapsnr}) to identify the highest (yellow) and lowest (dark blue) SNR values depending on the parameters $m_{1}$ and $m_{2}$ in each simulation and detector. From the three heatmaps we observe that the CE provides the highest SNRs, followed by NEMO and then ET. A common result to observe is that systems with primaries masses $m_1>1.52\;M_{\odot}$ give low SNRs. In summary, there is no apparent and uniform trend between both the masses and the signal intensity.

\begin{figure}
  \centering \includegraphics[width=\linewidth]{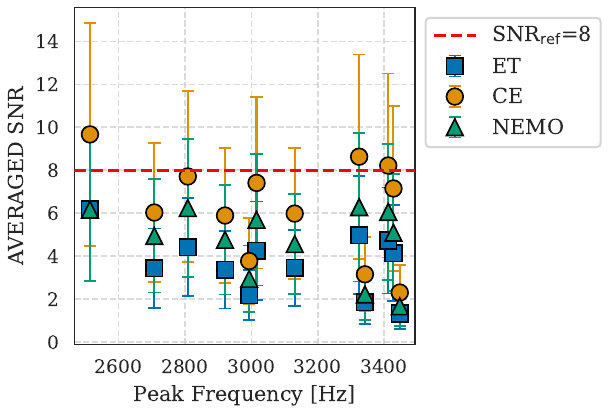}
  \vspace{0.5em} \includegraphics[width=\linewidth]{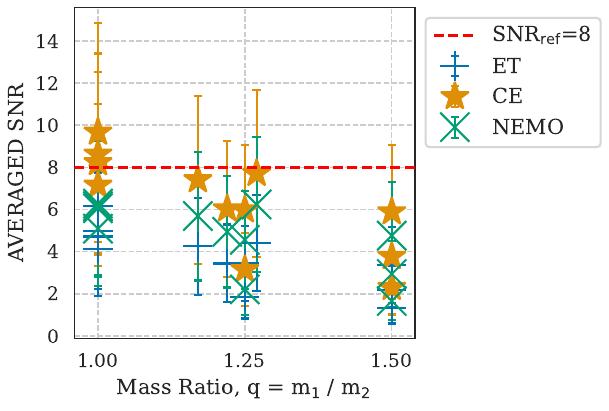}
  \caption{Detector sensitivity for Dataset 2. {\it Upper panel:} PM averaged SNR variation with the main peak frequencies of the characteristic strain. {\it Lower panel:} PM averaged SNR dependence on the mass ratio ($q = m_1/m_2$) of the inspiral system. Considering the central values, most of the simulations fall under the reference SNR.} 
  \label{fig:detectorSNRs_set2}
\end{figure}

In Fig.~\ref{fig:detectorSNRs_set2}, we see the distribution of the SNR according to the peak frequency and the mass ratio. In the upper panel, most of the PM signals are under the threshold line $\mathrm{SNR_{ref}}=8$. If we consider the central values of the averaged SNR, the performance of CE is $(9.67-6.17)/6.17\times 100\%=57\%$ better than that of ET and $(9.67-6.28)/6.28\times 100\%=54\%$ better than that of NEMO for the maximum SNR values. Even for minimum values, CE performs better with $73\%$ and $39\%$ over ET and NEMO, respectively. Based on the lower panel, we see the PM SNR and its distribution according to the inspiral mass ratio $(q)$. Equal-mass systems $(q\approx1.0)$ show higher SNRs than asymmetric coalescences $(q>1.17)$, especially for CE. For example, for BAM0125 ($q=1.0$) and BAM0127 ($q=1.5$), the CE SNR is $\sim 77\%$ greater for the first one. ET and NEMO follow a similar trend. This may suggest that mass-symmetric systems produce louder PM signals, probably due to more coherent oscillations in the remnant. For asymmetric systems, where less regular oscillation remnants may form, they emit higher frequency signals (e.g. $f_{2}\sim 3447\;\rm{Hz}$ for BAM0127) that fall into the less sensitive part of the detector. In summary, CE maintains a relative advantage even in asymmetric binary systems, therefore, it offers acceptable chances to observe the PM signals. 

In the following section, we explore the NS mass distribution based on electromagnetic observations and combine it with our SNR results to study the mass precision of M$_\mathrm{TOV}$.

\section{\label{sec:nsmasslimits}Neutron Star Mass Limit}
Mass measurements obtained through electromagnetic observations of pulsars in binary systems have revealed the type of distribution the neutron star mass follows, helping us to learn more about the formation channels of neutron stars in binary systems. These observations are essential in constraining the maximum mass that a neutron star can sustain before collapsing into a black hole. In this section, we first review the current neutron star mass distribution derived from electromagnetic observations and then discuss the precision for the Tolman-Oppenheimer-Volkoff mass (M$_{\mathrm{TOV}}$).

\subsection{\label{subsec:NSmassdistrib}Neutron Star Mass Distributions}
Studies based on the electromagnetic radiation emitted by pulsars have given some insights about the nature of ultra-dense matter in neutron stars, the astrophysical processes that generate the values of the estimated masses, and the evolutionary paths for the binaries in which they are in. These observations have provided a neutron star mass distribution that seems to follow a multimodal Gaussian, statistically favoring the bimodal and trimodal types \cite{OzelFreire2016, Antoniadis2016, Alsing2018, horvath2020birth}. This kind of distribution suggests that neutron stars do not arise from a single formation channel, as initially thought when the first observations were found to be around $1.4\;M_{\odot}$, because later neutron stars with lower masses around $1.25\;M_{\odot}$, and also with higher masses around $1.7-1.8\;M_{\odot}$, and even $2.01\;M_{\odot}$ were detected. The interpretation for each peak in the distribution is that $1.4\;M_{\odot}$ is attributed to the iron core collapse supernovae (FeCCSNe) process \cite{Timmes1996, Thorsett1999}, $1.25\;M_{\odot}$ is associated with electron capture supernovae (ECSNe) occurring in progenitors with O-Ne-Mg cores \cite{Nomoto1984, Podsiadlowski_2004}. The higher masses may be related to either accretion-induced mass growth in binary systems (like recycled pulsars) or a different population formed from more massive progenitors \cite{Kiziltan_2013, Tauris_2017, Fonseca_2021}. Additionally, these findings offer important constraints on the neutron star equation of state.

\begin{figure}
\includegraphics[width=\columnwidth]{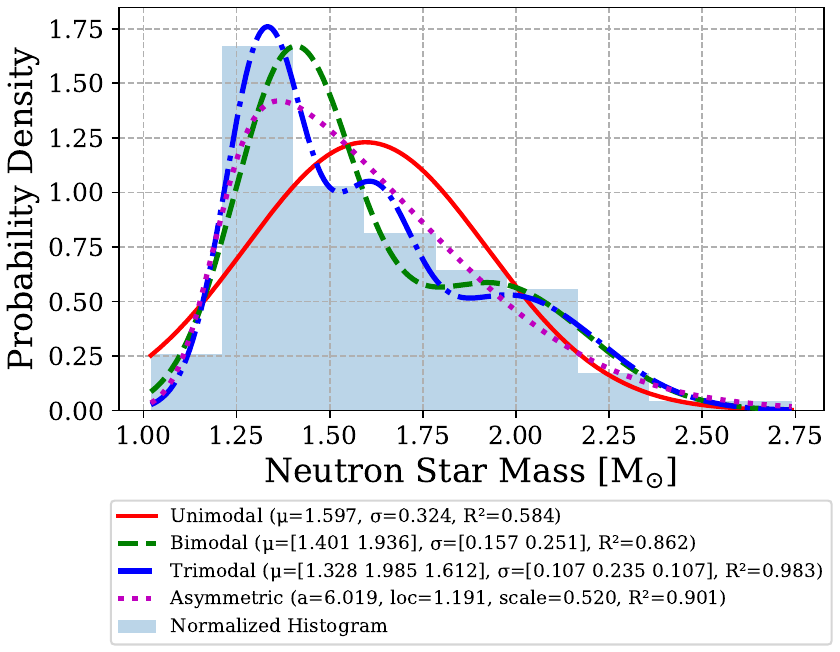}
\caption{Neutron star mass distribution for $122$ electromagnetic observations and four fitting Gaussian models: unimodal, bimodal, trimodal and asymmetric.}
\label{fig:122fitmassdistributions}
\end{figure}

Following the work of \cite{LucasdeSa2023} and \cite{Rochauniverse2023}, we use the same $122$ values for the mass data obtained from different binary systems, such as: X-ray binaries, binary neutron stars,  white dwarf-neutron star and massive star-neutron star. We test the fitting of unimodal, bimodal, trimodal, and asymmetric Gaussians to the mass distribution (see Fig.~\ref{fig:122fitmassdistributions}). The unimodal distribution is definitely ruled out not only because of the fit metrics but also because that distribution does not match all the observations obtained so far. The asymmetric Gaussian is not the best option either, but also it would be physically unclear, so we do not include it in our study. Then, we consider the bimodal distribution as the primary model because it has good-fit metrics and holds a theoretical support for each peak, as reported in previous papers. In future studies, we can consider the trimodal distribution due to theoretical/observational support for the three peaks and verify how this affects the mass precision in the $M_{\textrm{TOV}}$. 

\begin{figure}
\includegraphics[width=\columnwidth]{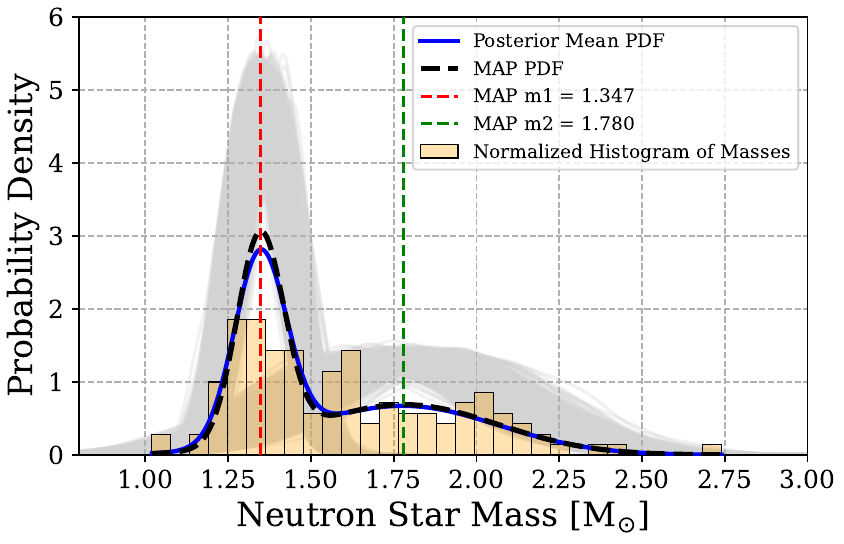}
\caption{Maximum a posteriori (MAP) probability density function (PDF) for 122 observed neutron star masses. The Bayesian estimation considers a bimodal gaussian likelihood and priors taken from Horvath et al., (2020).}
\label{fig:posterirmapbimodal122}
\end{figure}

The selection of the bimodal Gaussian leads to using it as the likelihood function in a Bayesian estimation for the neutron star mass distribution. For the priors, we used the information presented in \cite{horvath2020birth}, which is based on the observed properties and theoretical predictions for the neutron star mass distribution. The goal of using the Bayesian method is to obtain the estimate of the mass distribution by generating the Maximum A Posteriori Probability Density Function (MAP PDF) curve. This MAP PDF allows us to obtain the most probable value of the mass distribution given both a model and the observed data (see Fig.~\ref{fig:posterirmapbimodal122}). 

Regarding gravitational wave observations, the LIGO and Virgo detectors were able to estimate the masses of the merging objects in binary systems and provided information about the equation of state in the case of binary neutron stars. So far, two binary neutron star coalescences have been observed and named as: GW170817 and GW190425. The first detection occurred on August 17, 2017, where the masses of the neutron stars were estimated to be between $1.36\;M_{\odot}$ and $1.17\;M_{\odot}$ \cite{Abbott2017_GW170817}. This event was particularly special since after the emission of the gravitational radiation there were several electromagnetic follow ups such as a gamma-ray burst, a kilonova, and other posterior emissions in the EM spectrum. The analysis of the multimessenger detections allowed to constrain the equation of state of the neutron stars involved in the coalescence, finding that the softer EoS models were favored such as SLy and H4 over DD2 and APR \cite{Abbott_2018, LIGOScientific:2018hze}. In addition, the location and the luminosity distance of the source was found to be in the NGC4993 galaxy at $40\; \mathrm{Mpc}$ from Earth. The second detection occurred on April 25, 2019 and it involved neutron stars with masses of $1.12\;M_{\odot}$ and $2.52\;M_{\odot}$ \cite{LIGOScientific:2020aai}. In this case, no EM counterpart was observed, so the constraints on the EoS were less precise as well as with the luminosity distance to the source $(\sim159^{+69}_{-71}\;\mathrm{Mpc})$. The high mass range of the component masses suggest that neutron stars can be more massive than the typical $1.4\;M_{\odot}$, therefore, the mass limit can be extended to higher values. Nevertheless, the GW190425 studies do not discard the possibility of having a stellar black hole as the main component of the binary system. In the end, these two events, GW170817 and GW190425 are classified as the two BNS events so far detected through GW radiation. They were included in neutron star population studies in order to explore the mass distribution and still authors suggest that better constraints will be achieved by having more detections \cite{Landry_2021, Galaudage_2021}.  

We now focus on the total mass distribution of the binary system and the distribution of the remnant's mass. If the total mass is simply the sum of the masses of the two merging components $(M_{\mathrm{T}}=m_{1}+m_{2})$, then to obtain the distribution of the total mass, we apply the self-convolution to the obtained MAP PDF because when we add two independent distributions, the result is given by the convolution of the individual distributions. Mathematically, the convolution of two independent distributions $f_{1}(x)$ and $f_{2}(x)$ is defined as follows
\begin{equation}
(f_{1}*f_{2})(x)=\int_{-\infty}^{\infty}f_{1}(\tau)f_{2}(x-\tau)d\tau.
\end{equation}

To perform this calculation, we used the \texttt{numpy.convolve} function from Python's NumPy library to convolve the estimated MAP PDF. 

\begin{figure}
\includegraphics[width=\columnwidth]{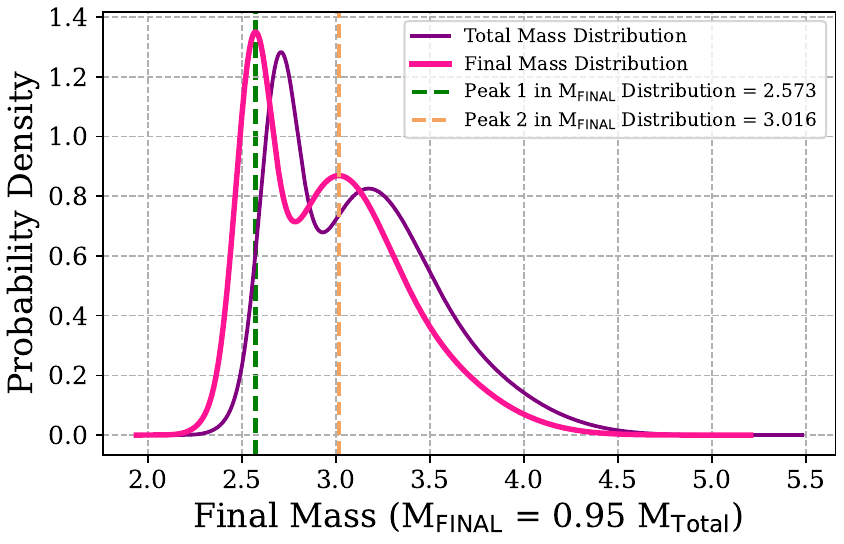}
\caption{The final mass distribution is the $95\%$ of the total mass, then this represents the remnant mass distribution for a bimodal estimation of neutron star masses in binary systems. The initial value of this distribution is $1.94\;M_{\odot}$ and the first peak is at $2.57\;M_{\odot}$.}
\label{fig:plotfinalalmass122}
\end{figure}

However, a portion of the mass is lost during the merger, mainly in the form of gravitational radiation. On the other hand, on the basis of simulations and theoretical models, the final mass ($M_{F}$) of the remnant can be reduced by approximately $5\%$ of the total mass, or in other words, the mass of the remnant is approximately $M_{F}=95\%\;M_{\mathrm{T}}$. This loss during the post-inspiral stages is typically due to winds of matter \cite{Fujibayashi_2018, Fujibayashi_2023}. This mass loss influences the final state of the remnant, which can be a neutron star if the mass is below the M$_{\mathrm{TOV}}$ limit or a black hole, otherwise. In Figure~\ref{fig:plotfinalalmass122} we see both the total mass and the final mass distributions. Clearly, the $5\%$ loss in the total mass distribution shows two peaks displaced to the left with respect to the total mass distribution. Then the main peaks are at $2.57\;M_{\odot}$ and $3.02\;M_{\odot}$. We also identify the initial value of the final mass distribution at $1.94\,M_{\odot}$. Here, we clarify that the pulsar mass observations are a starting point for the component masses in BNS systems, in which we assume that each neutron star is described by the PDF obtained from the observations. Then, when obtaining the total mass distribution, each PDF is independent but equal, which is why we use the self-convolution. Finally, since we are interested in the remnant (which could be a type of neutron star or a prompt collapse black hole) mass distribution, we applied the $5\%$ mass loss to the total mass distribution. This is how we are linking the observational pulsar data to the properties of the post-merger remnant. Certainly, this $5\%$ value has uncertainties, but it provides a reasonable more exact value for the main purpose of our work.

\subsection{\label{sec:mtovprecision}M$_{\mathrm{TOV}}$ mass precision from the PM stage}
Here, we merge the information obtained from the SNR values estimated in the matched filtering process with the final mass distribution curve.

For Dataset 1, we identify in Table~\ref{tab:Dataset1} the lowest and highest values for each detector. For example, for ET we obtain ${\mathrm{SNR_{low}}}=2.55\pm1.39$, corresponding to BAM0095, which uses the SLy EoS, and ${\mathrm{SNR_{high}}}=8.69\pm4.68$ for BAM0035, which uses the H4 EoS. In fact, these are the most favored EoS in the analysis performed for GW170817. Using the column of the SNR for ET, we also calculate the mean value of the SNR, resulting in ${\mathrm{SNR_{mean}}}=5.44\pm2.95$. Similarly, for CE the values are ${\mathrm{SNR_{low}}}=4.30\pm2.34$, ${\mathrm{SNR_{mean}}}=9.54\pm5.11$, and ${\mathrm{SNR_{high}}}=15.30\pm8.23$, while for NEMO they are ${\mathrm{SNR_{low}}}=3.15\pm1.72$, ${\mathrm{SNR_{mean}}}=7.62\pm4.07$, and ${\mathrm{SNR_{high}}}=12.85\pm6.90$.

If we follow the discussion presented in~\cite{Panther_2023}, then we assume $\rm{SNR_{ref}}\gtrsim 8$ as a valid reference signal-to-noise ratio for the PM stage. With this, we can determine the maximum distance at which the BNS signals we are analyzing would be detectable by future detectors. If the gravitational intensity is related to the distance from the source as $h\propto1/d$, then the maximum distance in terms of SNR is as follows
\begin{equation}
d_{m} = \frac{\mathrm{SNR_{i}}} {\mathrm{SNR_{ref}}}d,
\end{equation}

\noindent where $i=\mathrm{low, mean, high}$ in each detector and $d=40\;\mathrm{Mpc}$ as the distance at which the signals were extracted.  

Therefore, we find several values of $d_{m}$. Then we use these maximum distances to obtain the number of detections per year and in a spherical volume, $N$. We follow the definition and estimated values for the merger rate density in ~\cite{LIGOScientific:2025newR0}
\begin{equation}
R(z) = \frac{dN}{dV_{c} dt}(z),
\end{equation}

\noindent whose value at $z=0$ is in the range of $R_0=(7.6-250)\;\mathrm{Gpc^{-3}yr^{-1}}$. By integrating in one year and in a comoving volume $V_c=(4\pi/3) d_{\mathrm{m}}^{3}$, defined by the maximum distance, we obtain the number of detections $N$.

The results for the maximum distance are shown in Table~\ref{tab:dmax_snr}. We observe that the highest $d_m$ values come from the highest SNR for the three detectors: $[4.3\pm2.3, 7.7\pm4.1, 6.4\pm3.4]\;\rm{10^{-2}~Gpc}$.

\begin{table}
\caption{Maximum distance $d_\mathrm{m}$ (in Gpc) for Dataset~1. For each detector, the table lists the averaged SNR (low, mean, and high) together with their corresponding maximum distances, and all with their uncertainties.}
\resizebox{\columnwidth}{!}{%
\begin{tabular}{lcccccc}
\hline\hline
Detector & SNR$_\mathrm{low}$ & $d_\mathrm{m}^\mathrm{low}$ & SNR$_\mathrm{mean}$ & $d_\mathrm{m}^\mathrm{mean}$ & SNR$_\mathrm{high}$ & $d_\mathrm{m}^\mathrm{high}$ \\
   &   & $10^{-2}$~[Gpc] &   & $10^{-2}$~[Gpc]  &   & $10^{-2}$~[Gpc] \\ 
\hline
ET & $2.55\pm1.39$ & $1.3\pm0.7$ & $5.44\pm2.95$ & $2.7\pm1.5$ & $8.69\pm4.68$ & $4.3\pm2.3$ \\
CE & $4.30\pm2.34$ & $2.1\pm1.2$ & $9.54\pm5.11$ & $4.8\pm2.6$ & $15.30\pm8.23$ & $7.7\pm4.1$\\
NEMO & $3.15\pm1.72$ & $1.6\pm0.9$ & $7.62\pm4.07$ & $3.8\pm2.0$ & $12.85\pm6.90$ & $6.4\pm3.4$\\
\hline\hline
\end{tabular}
}
\label{tab:dmax_snr}
\end{table}

In table~\ref{tab:dN_detections} we summarize the results for the yearly number of detections. Here, we find that the most optimistic number of detections occurs when CE has the highest SNR, giving $N$: $0.47\pm0.76$ (maximum merger rate $R_0=250\;\mathrm{Gpc^{-3}yr^{-1}}$). 

\begin{table}
\caption{Estimated yearly number of BNS detections within a comoving spherical volume $V_c$ (radius $d_m$) derived from the SNR$_\mathrm{low,mean,high}$ values of each detector. For each case, $d_m$ is the maximum distance (in Gpc) used to compute the volume $V_c$ (in Gpc$^3$). Detection rates $N$ are then obtained for three representative merger-rate densities, 
$R_0=7.6,\;129,\;250$ Gpc$^{-3}\,\mathrm{yr}^{-1}$. Dataset 1.}
\resizebox{\columnwidth}{!}{%
\begin{tabular}{llccccc}
\hline\hline
SNR Case & Detector & d$_m$ & $V_c$ [Gpc$^{3}$] & N ($R_0$=7.6) & N ($R_0$=129) & N ($R_0$=250) \\
 & & $10^{-2}$~[Gpc] & & & & \\
\colrule
\multirow{3}{*}{Lowest} 
    & ET   & $1.3\pm0.7$ & $9\pm14\times$10$^{-6}$ & $0.7\pm1.1\times10^{-4}$ & $1.1\pm1.8\times10^{-3}$  & $2.2\pm3.5\times10^{-3}$ \\
    & CE   & $2.1\pm1.2$ & $4\pm7\times$10$^{-5}$ & $3.2\pm5.2\times10^{-4}$ & $5.4\pm8.7\times10^{-3}$  & $1.0\pm1.7\times10^{-2}$ \\
    & NEMO & $1.6\pm0.9$ & $2\pm3\times$10$^{-5}$ & $1.2\pm2.0\times10^{-4}$ & $2.1\pm3.4\times10^{-3}$  & $4.1\pm6.7\times10^{-3}$\\
\hline
\multirow{3}{*}{Mean}
    & ET   & $2.7\pm1.5$ & $8\pm14\times$10$^{-5}$ & $0.6\pm1.0\times10^{-3}$ & $1.1\pm1.8\times10^{-2}$ & $2.1\pm3.4\times10^{-2}$ \\
    & CE   & $4.8\pm2.6$ & $4\pm7\times$10$^{-4}$ & $3.4\pm5.6\times10^{-3}$ & $5.9\pm9.4\times10^{-2}$ & $1.1\pm1.8\times10^{-1}$ \\
    & NEMO & $3.8\pm2.0$ & $2\pm4\times$10$^{-4}$ & $1.8\pm2.8\times10^{-3}$ & $3.0\pm4.8\times10^{-2}$ & $5.7\pm9.3\times10^{-2}$\\
\hline
\multirow{3}{*}{Highest}
    & ET   & $4.3\pm2.3$ & $3\pm6\times$10$^{-4}$ & $2.6\pm4.2\times10^{-3}$ & $4.4\pm7.2\times10^{-2}$ & $0.9\pm1.4\times10^{-1}$ \\
    & CE   & $7.7\pm4.1$ & $2\pm3\times$10$^{-3}$ & $1.4\pm2.3\times10^{-2}$ & $2.4\pm3.9\times10^{-1}$ & $\mathbf{4.7\pm7.6\times10^{-1}}$ \\
    & NEMO & $6.4\pm3.4$ & $1\pm2\times$10$^{-3}$ & $0.8\pm1.4\times10^{-2}$ & $1.4\pm2.3\times10^{-1}$ & $2.8\pm4.5\times10^{-1}$ \\
\hline\hline
\end{tabular}
}
\label{tab:dN_detections}
\end{table}

Consequently, we estimate the uncertainty in the maximum neutron-star mass, $M_{\rm{TOV}}$, that future detectors could achieve from BNS post-merger signals using the empirical relation

\begin{equation}
\delta M (M_{F})= \frac{1}{N\; {\rm PDF}(M_{F})},
\label{eq:deltaM}
\end{equation}
\noindent where observationally we would say that the more detections we get, the better the precision, which is why $\delta M$ is inversely proportional to $N$, as well as to the normalized probability density function of the final mass distribution, because the most probable region, around $2.573 \,M_{\odot}$ in Fig.~\ref{fig:plotfinalalmass122}, leads to a better precision. 

To understand Eq. \ref{eq:deltaM} we may consider a uniform (flat) probability distribution for the final mass distribution between a lower limit $m_1$ and an upper limit $m_2$. If $N$ detections are made, the average separation between the highest mass neutron star and the lowest mass black hole would be $\delta M = (m_{2} - m_{1})/N$, this would be the uncertainty $\delta M$ in locating the limit value of $M_{\rm{TOV}}$. Since the normalized probability density for a uniform distribution is PDF = $1/(m_{2} - m_{1})$, this is equivalent to $\delta M= 1/(N\, \rm{PDF})$. So, Eq. \ref{eq:deltaM} generalizes this idea to any type of final mass distribution $\rm{PDF}(M_{F})$. Then, the precision $\delta M (M_{F})$ is mass-dependent of the remnant, that is, it is inversely proportional to the number of detections and the local probability density (PDF($M_{F}$)). Consequently, the uncertainty is smallest where the PDF is highest (the most probable remnant mass), and larger in the tails of the distribution. This follows from the intuitive notion that the threshold $M_{\rm{TOV}}$ can be constrained more precisely where the observed population is the densest.

\begin{figure}
\includegraphics[width=\columnwidth]{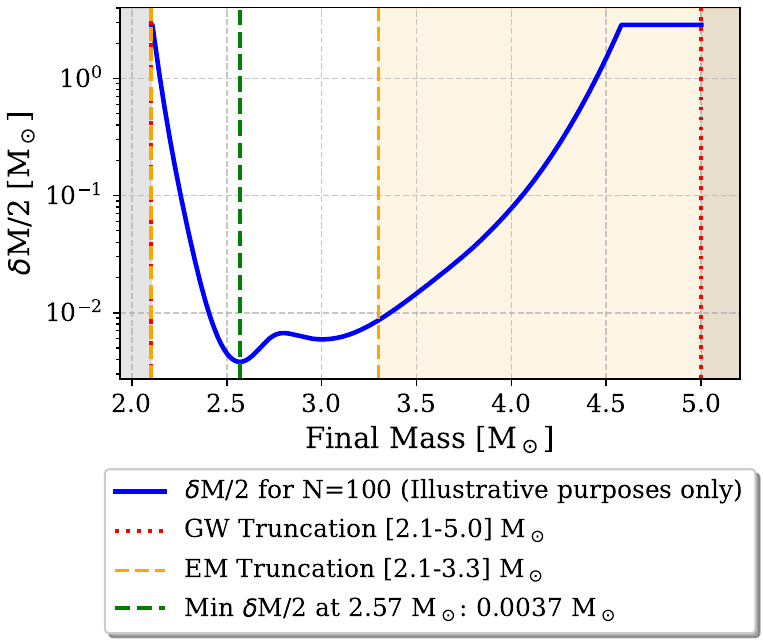}
\caption{Mass precision as a function of the final mass in the ranges $[2.1-5.0]\;\mathrm{M_{\odot}}$ and $[2.1-3.3]\;\mathrm{M_{\odot}}$, whose upper limits account for the commonly known mass gap and for the lowest mass stellar black hole detected. Considering a very optimistic situation in which the number of detections is $N=100$, the minimum precision is $0.37\%$. The choice of $N=100$ is for illustrative purposes only. The values obtained in our work are much lower, but the shape of the curve is similar.}
\label{fig:deltaM_Mfin_bi122_neweq}
\end{figure}

To have a comparative visualization of the shape of Eq.~\ref{eq:deltaM} and locate the minimum value, we first test an extremely optimistic and illustrative scenario with $N=100$ detections per year. Under this hypothetical case, the mass precision follows a curve with a minimum value at $2.57\;M_{\odot}$ and $\delta M/2=0.0037\;M_{\odot}$ (see Fig.~\ref{fig:deltaM_Mfin_bi122_neweq}). This value represents the minimum possible uncertainty since it is linked to the main peak (most probable) value of the final mass distribution, for any other point, the uncertainty would be larger, but in this work we provide the minimum value of $\delta M$. Therefore, it corresponds to expressing the maximum mass as $(\mathrm{M_{TOV}} \pm \delta M/2) \; M_{\odot} = \mathrm{M_{TOV}} \pm 0.037 \; M_{\odot}$. We emphasize that $N=100$ is purely for visualization, so in the mentioned figure, we show the mass precision distribution is truncated at two intervals $[2.1-5.0]\;M_{\odot}$ and $[2.1-3.3]\;M_{\odot}$, the maximum values being those referred to the mass gap upper limit $5.0\;M_{\odot}$~\cite{LIGOScientific:2024elc} and to the lowest stellar mass black hole observed $(3.3^{+2.8}_{-0.7}\;M_{\odot})$~\cite{doi:10.1126/science.aau4005}.

However, for the analysis of \textbf{Dataset 1}, we use the results obtained in Table~\ref{tab:dN_detections}, but choosing only those that hold $N\geq 1$. Then, for the three merger rates $R_0=[7.6, 129, 250]\;\mathrm{Gpc^{-3}yr^{-1}}$, we identify barely $1$ valid detection, giving the most optimal mass precision in the range $\mathbf{ \delta M/2 \approx 0.3-0.8}\;M_{\odot}$ for a merger rate upper bound $250\;\rm{Gpc^{-3}yr^{-1}}$ and the highest SNR of the Cosmic Explorer. It is important to clarify that a $0.8\,M_{\odot}$ precision means that it is the best possible uncertainty and corresponds to the main peak of the final mass distribution. With this, we note that the precision depends on the probability of each mass point, $\delta M (M_{F}) \approx 1 / (N\, \rm{PDF}(M_{F}))$.

For \textbf{Dataset 2}, we follow the same steps, but for the number of detections $N$ shown in Table~\ref{tab:Dataset2}. We identify the lowest, mean, and highest averaged SNR for each detector, obtaining $\mathrm{SNR_{low}}=1.33\pm0.74$, $\mathrm{SNR_{mean}}=3.69\pm1.98$, $\mathrm{SNR_{high}}=6.17\pm3.34$ for ET; $\mathrm{SNR_{low}}=2.30\pm1.28$, $\mathrm{SNR_{mean}}=6.33\pm3.38$, $\mathrm{SNR_{high}}=9.67\pm5.20$ for CE; and $\mathrm{SNR_{low}}=1.66\pm0.92$, $\mathrm{SNR_{mean}}=4.72\pm2.51$, $\mathrm{SNR_{high}}=6.28\pm3.46$ for NEMO. The results for the maximum distance are shown in Table~\ref{tab:dmax_snr_2} and the corresponding number of detections in Table~\ref{tab:dN_detections2}. As we see in the latter, no valid number of detections is obtained; therefore, we do not proceed to obtain the mass precision. 

\begin{table}
\caption{Maximum distance $d_\mathrm{m}$ (in Gpc) for Dataset 2. For each detector, the table lists the SNR values (low, mean, and high) together with their corresponding maximum distances and uncertainties.}
\resizebox{\columnwidth}{!}{%
\begin{tabular}{lcccccc}
\hline\hline
Detector & SNR$_\mathrm{low}$ & $d_\mathrm{m}^\mathrm{low}$ & SNR$_\mathrm{mean}$ & $d_\mathrm{m}^\mathrm{mean}$ & SNR$_\mathrm{high}$ & $d_\mathrm{m}^\mathrm{high}$ \\
& & $10^{-2}$ [Gpc] & & $10^{-2}$ [Gpc] & & $10^{-2}$ [Gpc]\\ 
\hline
ET   & $1.33\pm0.74$ & $0.7\pm0.4$ & $3.69\pm1.98$ & $1.9\pm1.0$ & $6.17\pm3.34$ & $3.1\pm1.7$\\
CE   & $2.30\pm1.28$ & $1.1\pm0.6$ & $6.33\pm3.38$ & $3.2\pm1.7$ & $9.67\pm5.20$ & $4.8\pm2.6$\\
NEMO & $1.66\pm0.92$ & $0.8\pm0.5$ & $4.72\pm2.51$ & $2.4\pm1.3$ & $6.28\pm3.46$ & $3.1\pm1.7$ \\
\hline\hline
\end{tabular}
}
\label{tab:dmax_snr_2}
\end{table}

\begin{table}
\caption{Estimated yearly number of BNS detections within a comoving spherical volume $V_c$ (radius $d_m$) derived from the SNR$_\mathrm{low,mean,high}$ values of each detector. For each case, $d_m$ is the maximum distance (in Gpc) used to compute the volume $V_c$ (in Gpc$^3$). Detections $N$ are then obtained for three representative merger-rate densities, 
$R_0=7.6,\;129,\;250$ Gpc$^{-3}\,\mathrm{yr}^{-1}$. These results correspond to Dataset 2.}
\resizebox{\columnwidth}{!}{%
\begin{tabular}{llccccc}
\hline\hline
SNR Case & Detector & d$_m$ & $V_c$ [Gpc$^3$] & N ($R_0$=7.6) & N ($R_0$=129) & N ($R_0$=250) \\
& & $10^{-2}$ [Gpc] & & & & \\
\colrule
\multirow{3}{*}{Lowest} 
    & ET   & $0.7\pm0.4$ & $1\pm2\times$10$^{-6}$ & $1.0\pm1.6\times10^{-5}$ & $1.6\pm2.7\times10^{-4}$ & $3.1\pm5.1\times10^{-4}$ \\
    & CE   & $1.1\pm0.6$ & $6\pm11\times$10$^{-6}$ & $4.8\pm8.1\times10^{-5}$ & $0.8\pm1.4\times10^{-3}$  & $1.6\pm2.7\times10^{-3}$\\
    & NEMO & $0.8\pm0.5$ & $2\pm4\times$10$^{-6}$ & $1.8\pm3.0\times10^{-5}$ & $3.1\pm5.1\times10^{-4}$ & $0.6\pm1.0\times10^{-3}$ \\
\hline
\multirow{3}{*}{Mean}
    & ET   & $1.9\pm1.0$ & $3\pm4\times$10$^{-5}$ & $2.0\pm3.2\times10^{-4}$ & $3.4\pm5.4\times10^{-3}$ & $0.7\pm1.0\times10^{-2}$\\
    & CE   & $3.2\pm1.7$ & $1\pm2\times$10$^{-4}$ & $1.0\pm1.6\times10^{-3}$ & $1.7\pm2.7\times10^{-2}$ & $3.3\pm5.3\times10^{-2}$\\
    & NEMO & $2.4\pm1.3$ & $5\pm9\times$10$^{-5}$ & $4.2\pm6.7\times10^{-4}$ & $0.7\pm1.1\times10^{-2}$ & $1.4\pm2.2\times10^{-2}$\\
\hline
\multirow{3}{*}{Highest}
    & ET   & $3.1\pm1.7$ & $1\pm2\times$10$^{-4}$ & $1.0\pm1.5\times10^{-3}$ & $1.6\pm2.6\times10^{-2}$ & $3.1\pm5.0\times10^{-2}$\\
    & CE   & $4.8\pm2.6$ & $5\pm8\times$10$^{-4}$ & $3.6\pm5.8\times10^{-3}$ & $0.6\pm1.0\times10^{-1}$ & $1.2\pm1.9\times10^{-1}$ \\
    & NEMO & $3.1\pm1.7$ & $1\pm2\times$10$^{-4}$ & $1.0\pm1.6\times10^{-3}$ & $1.7\pm2.8\times10^{-2}$ & $3.2\pm5.4\times10^{-2}$\\
\hline\hline
\end{tabular}
}
\label{tab:dN_detections2}
\end{table}

Compared to Dataset 1, the SNR values, the maximum distance values and the number of detections of Dataset 2 are lower, which might be related to the mass-asymmetry of the simulations in the second dataset.

The results for the mass precision on M$_{\rm{TOV}}$ for BNS remnant observations by future high-frequency band detectors are not optimistic unless in extreme situations of merger rate and SNR for the Cosmic Explorer. We point out that for the sake of generality of the method followed in this work, the results were obtained including cases of prompt-collapse black hole in the BNS simulations when the SNR values and therefore, the number of detections, were calculated.

\section{\label{sec:concperspec}Conclusions and perspectives}
Kilohertz gravitational-wave detectors such as ET, CE, and NEMO are expected to observe signals at high frequencies $(f>1\;\mathrm{kHz})$. This is interesting for the remnants produced in BNS events because their detection would provide information on the maximum mass of a neutron star. That is why we focus on assessing the mass precision of $M_{\mathrm{TOV}}$ for these detectors using the post-merger signals of BNS simulations. For this purpose, we used two sets of simulations from the CoRe database. The first dataset is characterized for having neutron stars with equal masses and varying only the equation of state in each simulation, while the second dataset varied the neutron star masses and fixed the EoS to SLy. Both datasets offered waveforms that capture important characteristics of BNS coalescences. 

Focusing on the PM stage, we extracted this portion of the strain signals and computed their amplitude spectral densities (ASDs) to identify the main peak frequencies. We then derived the corresponding characteristic strains to evaluate the detectability of these signals by next-generation detectors such as ET, CE, and NEMO. For the sake of generality in the source sky position, we averaged the SNR over 4 angles via MCMC and assumed that the binary neutron stars are located at a distance of $40\;\mathrm{Mpc}$, based on the estimations of GW170817. From the resulting SNR values, we identified the lowest ($\mathrm{SNR}_{\mathrm{low}}$), highest ($\mathrm{SNR}_{\mathrm{high}}$), and mean ($\mathrm{SNR}_{\mathrm{mean}}$) values for each detector.

To estimate the maximum distance $d_m$ at which each signal remains detectable, we used the proportionality between the strain amplitude and the distance $(h\propto1/d)$, rescaling the reference distance according to the ratio between each SNR and a detection threshold of $\mathrm{SNR_{ref}} = 8$. Using these maximum distances, we computed the expected number of binary neutron star detections per year for each detector by evaluating the comoving volume enclosed within each detection range and combining it with three reference merger rate densities. Adopting representative values of $R_0 = [7.6, 129, 250]\; \mathrm{Gpc}^{-3}\, \mathrm{yr}^{-1}$ and assuming one year of observation time, we obtained the corresponding detection numbers.

These values were then used to estimate the precision with which the neutron star mass distribution can be constrained. Specifically, for every valid number of detections ($N\geq1$), we computed the corresponding mass uncertainty $\delta M/2$.

To do this, we needed to work with the final mass distribution of neutron stars that are in binary systems. For this matter, we used observational data obtained from electromagnetic sources and fit Gaussian models. From this we determined that a bimodal mass distribution fits well with the observed data, with a trimodal Gaussian model also providing an acceptable alternative to be performed in posterior works. The fit helps us define the likelihood for the Bayesian analysis, and the priors were taken from previous work on neutron star mass distributions. The Bayesian inference refines the mass distribution estimates by obtaining the Maximum A Posteriori Probability Density Function curve. This allowed us to obtain the total mass distribution of the binary systems by applying a self-convolution to the MAP PDF data, considering that they are independent distributions. From that result, the most probable values were found to be $2.71\;M_{\odot}$ and $3.18\;M_{\odot}$ for the total mass. Subsequently, we considered the mass loss ($5\%$) after the merger in order to obtain the distribution of the final mass. Even if this value carries uncertainties from previous works, it is sufficiently true for the main focus of our work. Then, the final mass distribution is shifted, and the peaks at $2.57\;M_{\odot}$ and $3.02\;M_{\odot}$ are consistently in the range of known events such as GW190425 and GW230529~\cite{LIGOScientific:2024elc}, being the estimated masses $3.6^{+0.8}_{-1.2}\;M_{\odot}$ and $1.4^{+0.6}_{-0.2}\;M_{\odot}$ in the latter event. 

From the results, only Dataset 1 provides marginally one valid number of detections. The lowest and mean merger rates are insufficient to obtain a reasonable $N$. If we focus on the most optimal scenario, we have that for \textbf{Dataset 1}, the highest SNR of CE is: $15.30\pm8.23$, for which the highest valid number of detections is $N=0.47\pm0.76$ for the maximum merger rate $250\;\mathrm{Gpc}^{-3}\mathrm{yr}^{-1}$. Then, the mass precision is between $\mathbf{\delta M/2 \approx 0.3-0.8\;M_{\odot}}$ under the most optimistic scenario for SNR and merger rate. This range corresponds to the central and upper limit values of $N$, since the lower limit is negative and thus, unphysically valid.

We emphasize that this value is the minimum uncertainty that corresponds to the most probable mass in the final mass distribution (main peak or highest density in the PDF). Then, for other final masses, the uncertainty would be even larger, as it is an inverse function of the PDF($M_{F}$).

On the other hand, the results are worse for \textbf{Dataset 2}, giving no valid $N$ and meaning that the mass-asymmetry binaries provide PM signals that are more challenging to observe. 

Based on the results we got from this study, to have better BNS PM detections in next-generation earth-based detectors, it will still be necessary to improve their sensitivity at high frequency. 

\section*{Acknowledgements}

This study is part of the PhD Program at the Astrophysics
Division of Instituto Nacional de Pesquisas Espaciais, São José dos Campos, Brazil. Gabriela Conde Saavedra is deeply grateful for the ﬁnancial support given by the Coordenação de Aperfeiçoamento de Pessoal de Nível Superior - Brasil (CAPES) Finance Code 001. H. P. de Oliveira thanks Conselho Nacional de Desenvolvimento Cient\'ifico e Tecnol\'ogico (CNPq) and Funda\c c\~ao Carlos Chagas Filho de Amparo \`a Pesquisa do Estado do Rio de Janeiro (FAPERJ)
(Grant No. E-26/200.774/2023 Bolsas de Bancada de Projetos (BBP)). ODA thanks the Brazilian Ministry of Science, Technology and Innovation and
the Brazilian Space Agency (AEB), which supported the present work under PO 20VB.0009. He also thanks CNPq for grant number 310087/2021-0. MU thanks CNPq for the Bolsa de Produtividade em Pesquisa (PQ), process number 302954/2025-2.

G. Conde-Saavedra is grateful to J. M. S. de Souza for providing the matched-filtering code and giving feedback to preliminary results, as well as to L. M. de Sá, L. S. Rocha and J. E. Horvath for sharing and answering questions about the neutron star mass data distribution. G. C.-S. is also thankful to the gravitational wave group at the Instituto Nacional de Pesquisas Espaciais (GWINPE) for the discussions during the weekly meetings, and to the LIGO/CBC/Extreme Matter Group for the valuable suggestions. 

\bibliographystyle{apsrev4-2}
\bibliography{Paper1_deltaM_PRD} 
\end{document}